\documentclass{icrc29}
\usepackage{graphicx,amssymb,amsmath,times}
\begin{document}

\title{\bf {Fluorescence and Hybrid Detection Aperture of the Pierre Auger Observatory}}

\author[J.A. Bellido et al.]{J.A. Bellido, D. D'Urso, H. Geenen, F. Guarino, L. Perrone, S. Petrera, L. Prado Jr.,
\newauthor F. Salamida for the Pierre Auger Collaboration}

\presenter{Presenter:S.Petrera (sergio.petrera@aquila.infn.it), \  
ita-petrera-S-abs1-he15-poster}

\maketitle

\begin{abstract}
The aperture of the Fluorescence Detector (FD) of the Pierre Auger Observatory is 
evaluated from simulated events using different detector configurations: mono, stereo, 
3-FD and 4-FD. The trigger efficiency has been modeled using shower profiles with ground 
impacts in the field of view of a single telescope and studying the trigger response (at 
the different levels) by that telescope and by its neighbours. In addition, analysis cuts imposed by 
event reconstruction have been applied.
The hybrid aperture is then derived for the Auger final extension.
Taking into account the actual Surface Detector (SD) array configuration and its trigger 
response, the aperture is also calculated for a typical configuration of the present phase.
\end{abstract}

\section{Introduction}
\label{sec:intro}

One of the main objectives of the Pierre Auger Observatory \cite{EA} is the measurement of the flux of cosmic
rays above 10$^{18}$ eV. An accurate knowledge of the aperture is an important piece of this measurement, in particular
in its different detection configurations: SD-only, FD-only and hybrid. For 
this purpose it is mandatory to use a detailed simulation to exploit dependences on the numerous parameters.

The detector aperture, integrated over solid angle, is given by
\begin{equation}
{\mathcal{A}} = \int_\Omega S_{eff} ~ cos\theta ~d\Omega \quad ,\quad \mathrm{ with} \quad \quad
S_{eff} = \int_{A_{gen}} \epsilon_{trg} \cdot \epsilon_{rec}  ~d S
\label{apert}
\end{equation}
where $d\Omega = d cos \theta d\phi$ and $\Omega$ are the differential and total solid angles 
respectively. Excluding inclined showers, 
$\theta$ varies from 0 to $60^{\circ}$. $ S_{eff}$ is the effective area,
$\epsilon_{trg}$ is the overall trigger efficiency and $\epsilon_{rec}$ is the reconstruction
efficiency. $d S$ and $A_{gen}$ are
the differential and total areas where shower events hit ground level. 
Equation (\ref{apert})  allows the calculation of aperture for different conditions and
detector configurations. Single detector (SD or FD) apertures are obtained using the relevant detector efficiency functions for 
$\epsilon_{trg}$ and $\epsilon_{rec}$. On the other 
hand, a hybrid aperture calculation
has to include efficiencies of both detectors and therefore each efficiency function 
has to be read in
(\ref{apert}) as the product $\epsilon^{SD} \times \epsilon^{FD}$ \footnote{
Hereafter, the upper index will be explicitly written only in case of ambiguity, when both detections
are considered at the same time.}.

\section{The basic FD trigger function}
\label{sec:basf}

Under the assumption that all six telescopes forming an FD eye 
have identical characteristics one can restrict the generation area 
 to a slice with a vertex at a telescope, radius from 0 up to a maximum distance
and azimuthal width equal to the telescope field of view (i.e. $\pm 15^{\circ}$ around the telescope axis).
Efficiency is calculated as a function of polar coordinates of shower impact points. 
Because of left-right symmetry, efficiency is a function of $r$ and 
 $\Phi = |\phi_G - \phi_G^{axis}|$,
 the absolute azimuthal distance of the impact point from the telescope axis azimuth. 
The assumption of identical telescopes is needed to extend the evaluation of $\epsilon_{trg}(r,\Phi)$ for $\Phi > 15^\circ$, 
using the efficiency measured by the neighbouring telescopes.
 %\cite{Ap1}.
The simulation FDSim \cite{NIM} was used for generating showers from two primary types, protons and iron nuclei, 
in 8 energy bins  from 10$^{17}$ to 10$^{20.5}$ eV.
Two different atmospheres were used with  vertical aerosol optical depth (VAOD) 
above the FD level of 0.03 and 0.06, respectively. These atmospheres will be referred to as ``clean'' and
``dirty'', having respectively lower and higher VAOD with respect to the typical Malarg\"ue atmosphere\cite{Atm}.

In total 700,000 shower profiles have been generated. Trigger efficiency has been calculated in cells of $\Delta r = 2$ km times $\Delta \Phi = 1.5^\circ$ for
four separate cases: protons and iron nuclei in ``clean'' and ``dirty'' atmospheres. The basic trigger function 
$\epsilon_{trg}(r,\Phi)$ has
been regularized and parameterised, at each energy, through a fit to an empirical function.

During the current period of data taking only some telescopes are fully equipped with a
corrector ring (an optical device for increasing aperture) \cite{EA}. 
Therefore, for a selected configuration of primary and atmosphere (protons in ``clean'' atmosphere), we simulated 
20,000 shower profiles with this optics, to study the differences 
between telescopes with or without a corrector ring.

\section{The hybrid aperture in ideal case}
\label{sec:pshyb}

Using the basic trigger function discussed in the previous section 
it is straightforward to evaluate the efficiency  $\epsilon_{trg}^{OR}$ to trigger a shower by 
{\it at least one telescope in the eye}.~ 
This function is mapped onto the ground.
For this purpose the ground area
is divided into a grid of 0.5 $\times$ 0.5 km$^2$ cells and 
the corresponding polar coordinates $(r,\phi_G)_{eye}$ relative to each eye is calculated.
This allows the evaluation of $\epsilon_{trg}^{OR}$ at the cell center as seen by each eye
for the different detection configurations, i.e.\ {\it mono}, {\it stereo}, {\it 3-FD} or {\it 4-FD}. 
%In Fig. \ref{Efix} trigger efficiencies are shown at $Log(E/eV)$ = 19.5 for the different FD detection configurations.
The corresponding aperture
follows directly from the surface integral in eq. (\ref{apert}). The integral has been performed over 
a domain fully including the SD array in its final extension and assuming 
$\epsilon_{trg}^{SD} = 1$ within the array boundary and 0 outside. This represents
the {\it effective hybrid area in the ideal case of a fully efficient SD at any energy}. 
Since a hybrid trigger requires only a SD single station trigger (SD-T1)\cite{EA}, this condition
turns out to be fulfilled as soon as the energy exceeds 10$^{18}$ eV. 

Then the aperture is $ {\mathcal{A}} = \bar{S}_{eff} \times \pi (1 - cos^2 \theta_{max}) \simeq 2.36 ~sr \times  \bar{S}_{eff},$
where the bar signs stand for the average over solid angle,
with a $cos\theta$ $d cos\theta$ distribution.
%%%%%%%%%%%%%%%%%%%%%%%%%%%%%%%%%%%%%%%%%%%%%%%%%%%%%%%%%%%%%%%%%%%%%%%%%%%%
\begin{figure}[t]
\centering
\begin{tabular}{p{.45\textwidth} p{.45\textwidth}}
\centering
\includegraphics[width=.45\textwidth]{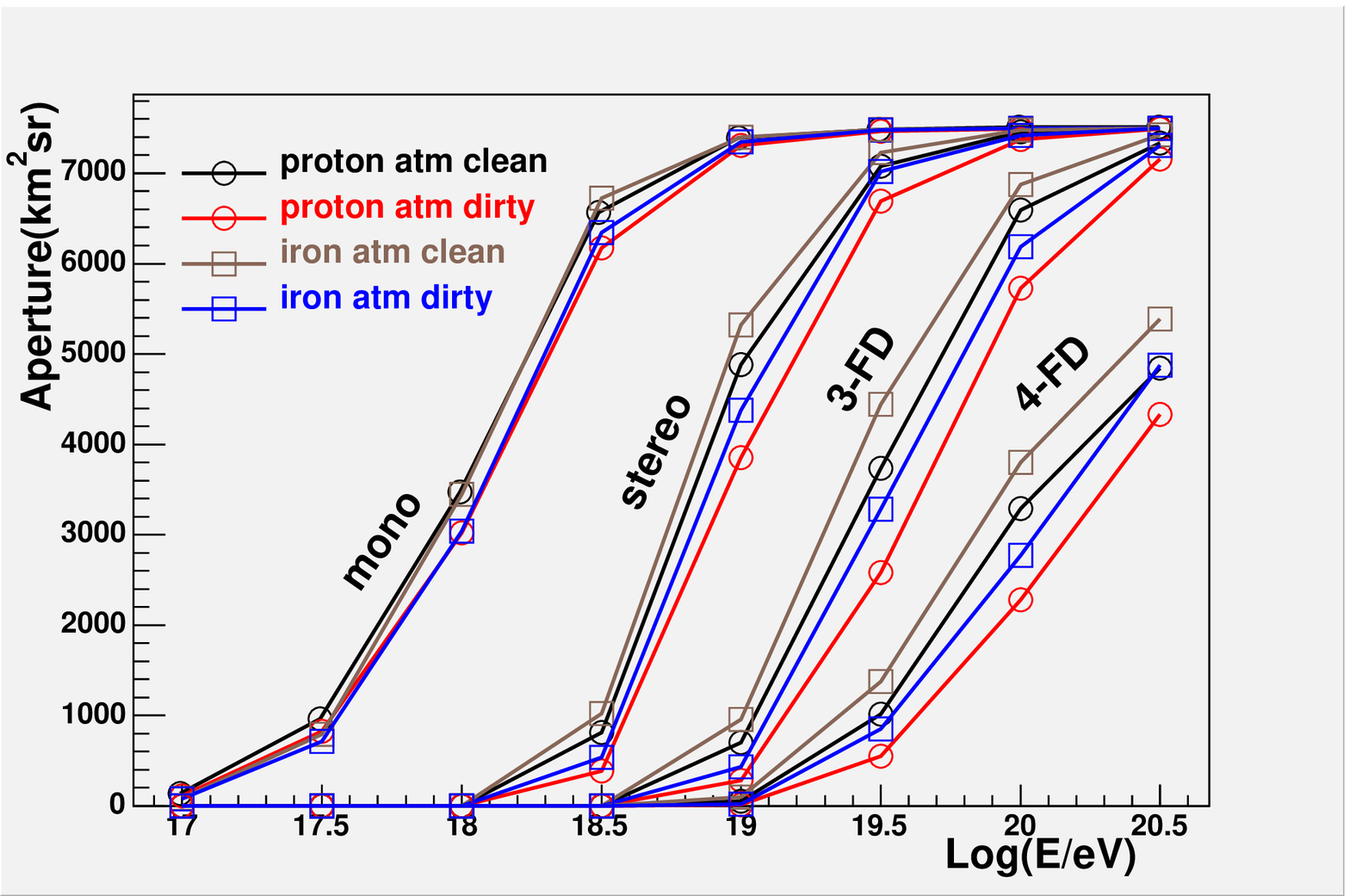} \caption {\small{Estimated hybrid apertures in 
km$^{2}\cdot$sr as a function of shower energy for different detector configurations, primary particles and atmospheric conditions.}}\label{hyb}
&
\includegraphics[width=.45\textwidth]{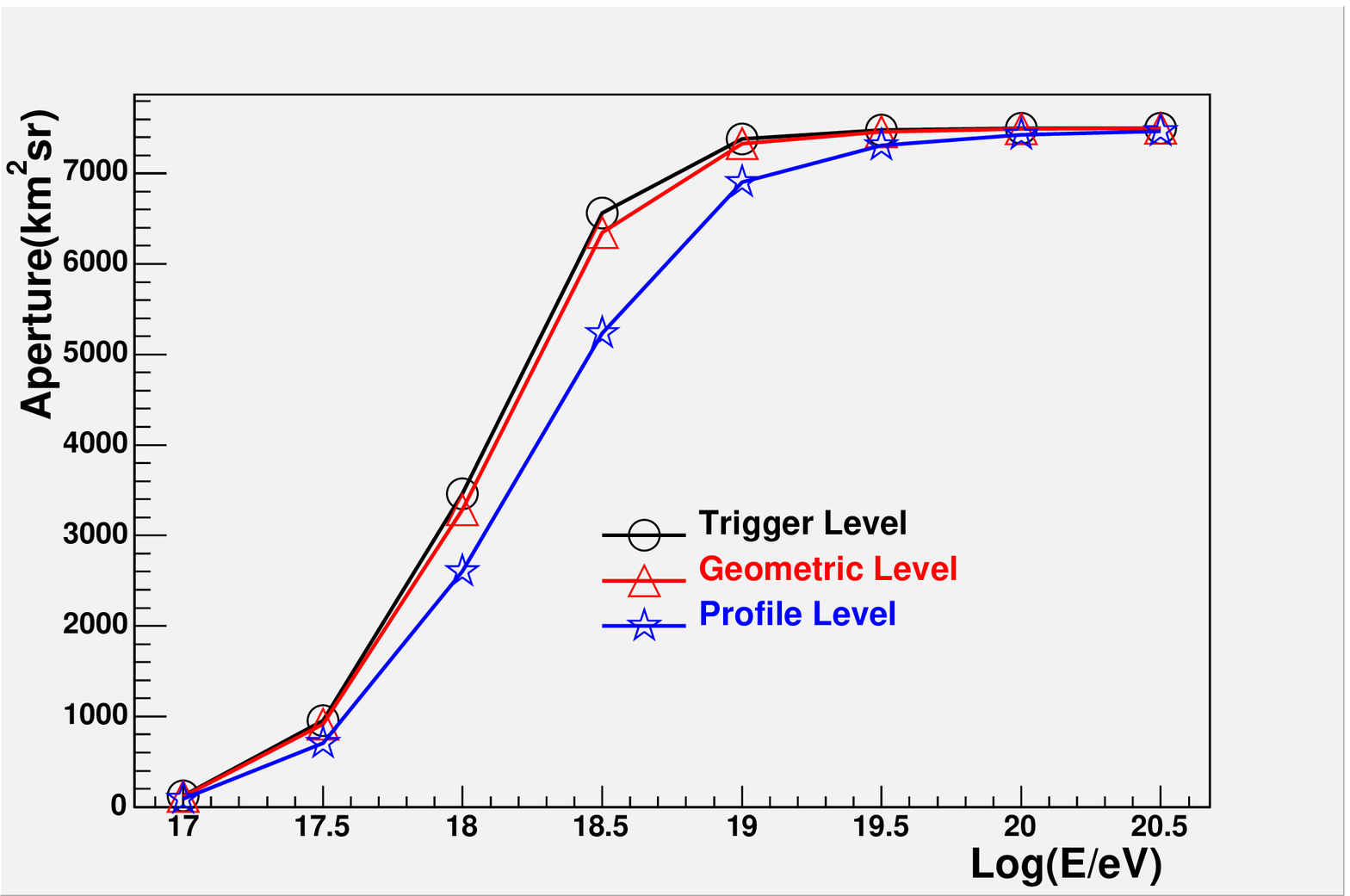}\caption{ \label{rec} \small {Estimated mono hybrid apertures in km$^{2}\cdot$sr at various reconstruction levels as a function of shower energy for proton primaries in a ``clean'' atmosphere.}}
\end{tabular}
\end{figure} 
%%%%%%%%%%%%%%%%%%%%%%%%%%%%%%%%%%%%%%%%%%%%%%%%%%%%%%%%%%%%%%%%%%%%%%%%%%%%
In Fig. \ref{hyb} the aperture calculated for different FD configurations is shown. 
The mono-hybrid aperture is practically independent of primary type, but
is slightly sensitive to atmosphere parameters. The relative difference between the two atmospheres 
is about 15\%  at $10^{18}$ eV and vanishes at 
higher energies. For more demanding detection configurations (stereo, 3-FD and 4-FD), apertures tend to be more sensitive to both
composition and atmosphere.

Apertures shown in Fig. \ref{hyb} are trigger apertures. since the only requirement applied to showers is
the fullfillment of trigger conditions. Other conditions have to be applied if showers are to be reconstructed.
In Fig. \ref{rec} we show how the aperture changes when the Shower Detector Plane (SDP) ({\it Geometry Level}) and
the shower profile ({\it Profile Level}) are reconstructed. In this figure, SDP reconstruction was assumed
to be fulfilled with at least 5 pixels. The profile level requires a minimum number of 10 pixels and the shower
maximum in the telescope field of view.

\section{The hybrid aperture of the growing detector}
\label{sec:acthyb}

The aperture shown so far is based on FD efficiency only and for
the full Auger site as the generation area. 
However the Auger detector
is not static, but is continually changing:
%\begin{itemize}
at present, since new tanks are added and put into operation 
and, at completion, since individual detectors (SD tanks and/or FD telescopes) can be occasionally 
out of operation or excluded because of non standard performance.
%\end{itemize}
Therefore the actual hybrid aperture in the construction phase and also in the final configuration 
has to be calculated
using both FD and SD trigger responses.
The SD trigger efficiency has been extensively studied separately 
%within the ``SD trigger and acceptance'' working
%group. 
and in particular the use of a ``Lateral Trigger Probability'' (LTP) function has been proposed \cite{Par1} for
 a complete description of the trigger features. 
These LTP functions have been  parameterised as a function of distance at various energies. 
\begin{figure}[b]
\centering
\includegraphics[width=9.8cm]{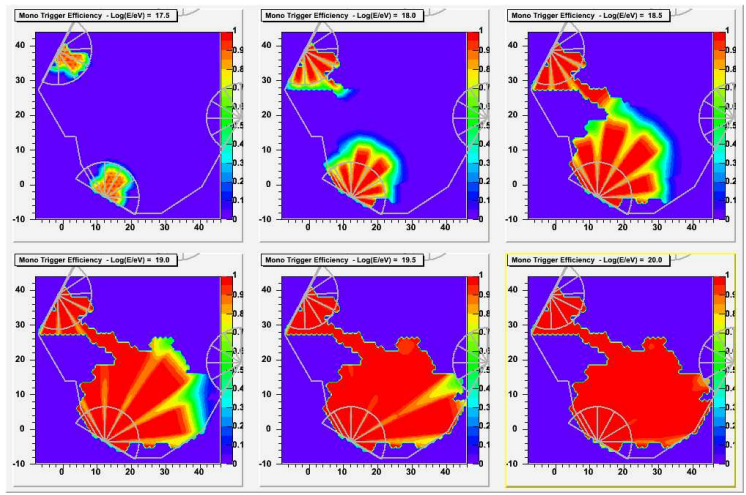}\caption {\small{Mono hybrid trigger efficiency 
at mid October 2004, for proton 
primaries in ``clean'' atmosphere, at energies from $10^{17.5}$ to $10^{20}$ eV. 
Efficiency is represented following a
color code from violet (0) to red (1). At this date, the two central 
telescopes (3 and 4) of both eyes were equipped with corrector ring lenses. The unusual shape
of the domain where efficiency is greater than 0  is caused by the extent of the SD array at this 
date.
}}\label{hmono}
\end{figure} 

\begin{figure}[t]
\centering
\begin{tabular}{p{.45\textwidth} p{.50\textwidth}}
\centering
\includegraphics[width=.45\textwidth]{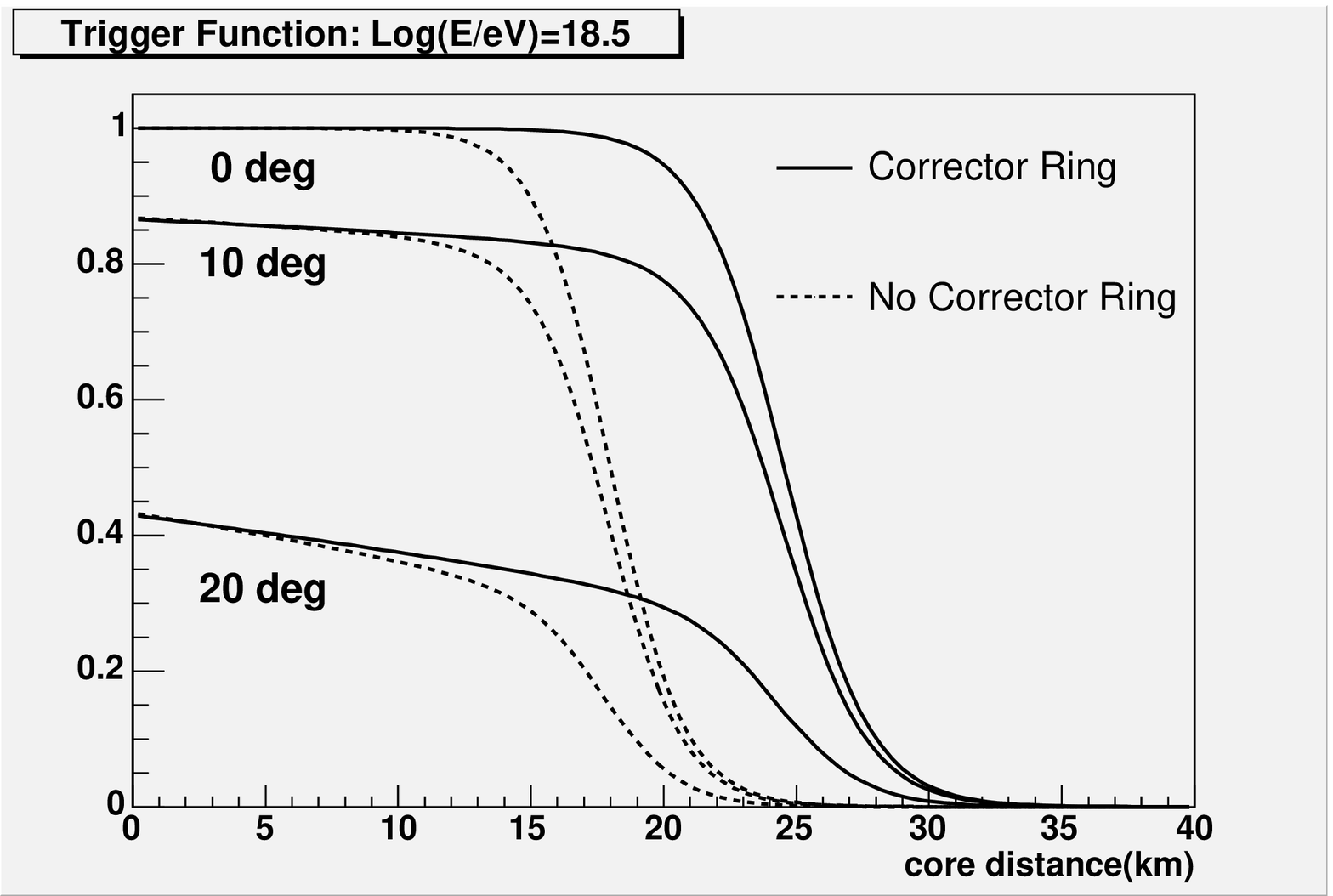}\caption {\small{ FD basic trigger function 
$\epsilon_{trg}(r,\Phi)$ at  10$^{18.5}$ eV for three different $\Phi$ angles. 
The function parameters refer to proton primaries in a ``clean'' atmosphere.}}\label{crNocr} &
\includegraphics[width=.50\textwidth]{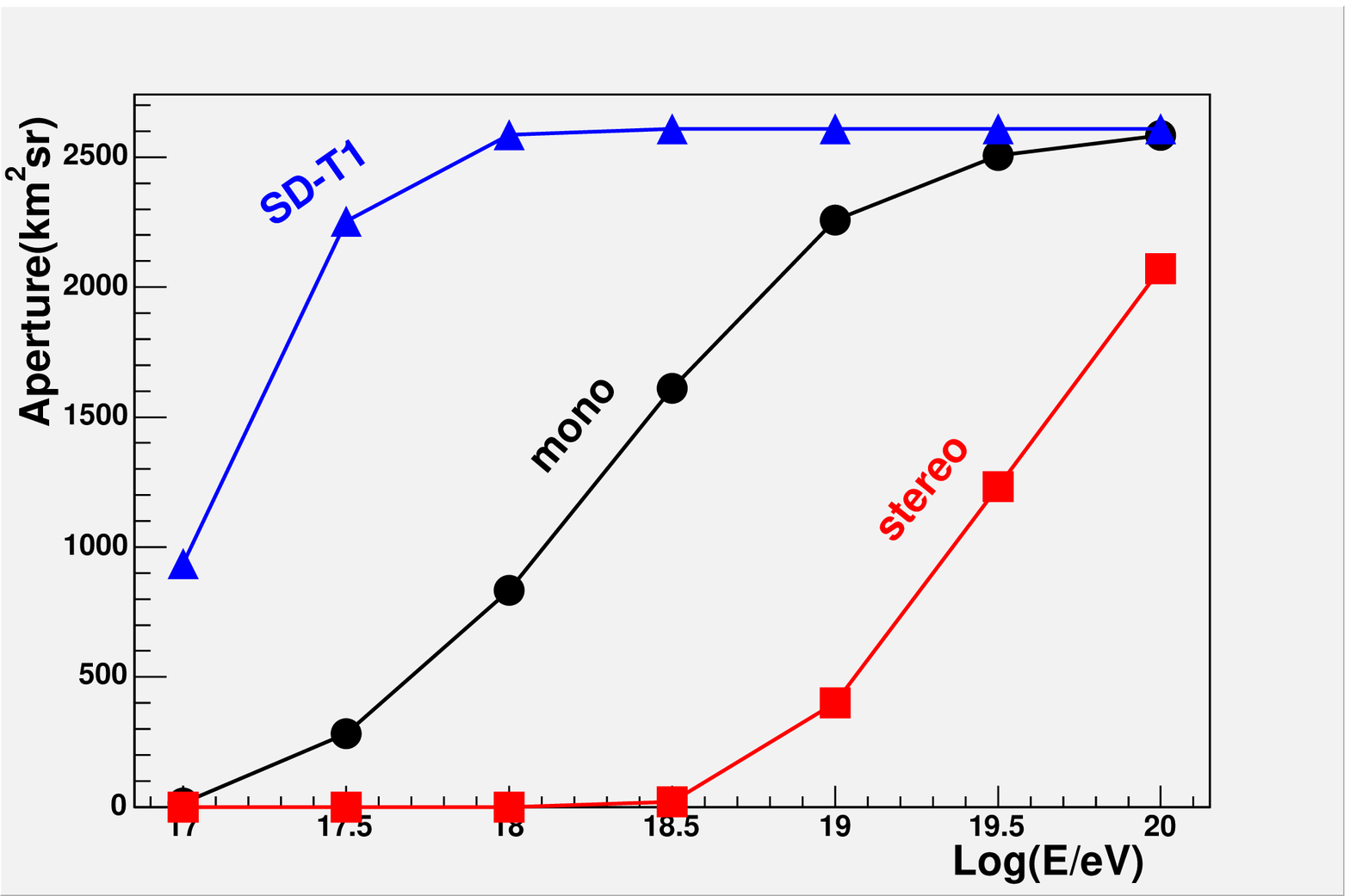}\caption {\small{Hybrid Trigger aperture in 
km$^{2}\cdot$sr for mono, stereo and SD-T1 detections.}}\label{aeffNow}
\end{tabular}
\end{figure}

Because in the current phase of data taking FD telescopes have different optical configurations (with and without 
corrector ring),
the basic trigger functions were calculated accordingly.
As shown in  fig \ref{crNocr},
the configuration with corrector ring allows more distant viewing at any angle.

In Fig. \ref{hmono} the mono hybrid trigger 
efficiency is shown  as a function of energy, for the site configuration of 
mid October 2004. %The interplay between SD and FD can be easily recognized. 
At lower energies the mono hybrid trigger accesses only the area covered by SD tanks just in front of each eye.
In particular at $10^{17.5}$ eV the single tank are easily visible as red spots in the trigger 
patterns. At higher energies, when the LTP range exceeds the distance between contiguous tanks, the area accessed
by the hybrid trigger is roughly the intersection of FD patterns with SD domain. 
%New regions are further covered mainly because of merging of different telescopes and different eyes. 

%%%%%%%%%%%%%%%%%%%%%%%%%%%%%%%%%%%%%%%%%%%%%%%%%%%%%%%%%%%%%%%%%%%%%%%%%%%%
In Fig. \ref{aeffNow} the hybrid trigger aperture is shown as a function of energy for mono
and stereo detection as compared with SD-T1 detection\footnote
{
SD data are acquired at higher trigger level (namely T3) since trigger rates are too high at lower levels.
} 
(SD-only single station trigger efficiency). 
It can be seen that mono
detection approaches SD-T1 as energy increases. Stereo detection becomes
sizable above $10^{19}$ eV and then rapidly increases. 

\section{Conclusions}
\label{sec:concl}

The use of basic trigger functions for both SD and FD detectors allows a detailed study of the
local trigger probability and aperture in the case of hybrid detection.
This approach is particularly appealing since it limits the intensive CPU simulation to the
generation of these functions, that can then be used very easily for aperture calculations.
The advantage is still more evident if one takes into account that simulation studies
are largely detector specific.
Finally we have shown that this approach allows the continuous evaluation of apertures
as new detectors are added in the field and put into operation.

\end{document}